\documentclass[aps,showpacs]{revtex4}
\begin{document}
\title{No news for Kerr-Schild fields}
\author{Boyko V. Ivanov}
\email{boyko@inrne.bas.bg}
\affiliation{Institute for Nuclear Research and Nuclear Energy,\\
Tzarigradsko Shausse 72, Sofia 1784, Bulgaria}

\begin{abstract}
Algebraically special fields with no gravitational radiation are described.
Kerr-Schild fields, which include as a concrete case the Kinnersley photon
rocket, form an important subclass of them.
\end{abstract}

\maketitle

\section{Introduction}

In 1951 Vaidya \cite{one} generalized the Schwarzschild vacuum solution to a
spherically symmetric ''shining star'' solution, emitting pure radiation. In
1969 Kinnersley \cite{two} further generalized this solution to a photon
rocket - a particle of arbitrary acceleration emitting an anisotropic jet of
radiation. He noted that it belonged to the Kerr-Schild (KS) class of
metrics, studied till then in vacuum \cite{three} and in the presence of
electromagnetic fields \cite{four}. Later it was shown that the photon
rocket is also a type D pure radiation solution from the Robinson-Trautman
(RT) class (algebraically special expanding solutions without a twist
(rotation)) \cite{five}. In 1994 Bonnor \cite{six} found that axially
symmetric photon rockets do not emit gravitational radiation, which is
strange for an accelerating particle. A suggestion was made, based on the
linearised theory, that axial symmetry was responsible for this phenomenon 
\cite{seven}. However, an axially symmetric RT solution of type II was
studied as a anisotropic perturbation of the photon rocket, still with no
sign of gravitational radiation \cite{eight}. Another explanation was given 
\cite{nine}, based on the well-known vanishing of the Einstein
pseudo-energy-momentum tensor for KS fields \cite{ten}. It was extended to
the Einstein-Maxwell field of a charged particle, undergoing arbitrarily
accelerated motion \cite{eleven}.

The fact that the photon rocket is a member of the RT class was emphasized
in \cite{twelve,thirt}. In the latter reference the Bondi-Sachs \cite
{fourt,fift} formalism was used and applied to axially symmetric RT pure
radiation solutions. It was shown that the Kinnersley solution is the only
one which does not radiate gravitationally. This put the study of photon
rockets on firm basis, since different pseudo-tensors for KS metrics lead to
different results \cite{sixt}. Later the news function was calculated for
any RT metric and it was found once more that the Kinnersley solution is the
only one to emit just photon radiation \cite{sevent,eightt}. Rotating pure
radiation KS fields (radiating Kerr metric) were given in \cite{ninet,twenty}
and it was demonstrated \cite{twenty} that only pure radiation contributes
to the mass loss.

Finally, the Bondi mass and the news function were recently found for
twisting algebraically special metrics in a paper \cite{twone}, based on
previous research of Tafel and co-authors \cite{twtwo,twthree}, where the RT
class was discussed as a special case.

Meanwhile, all pure radiation KS fields with axial symmetry have been found
explicitly \cite{twfour}. Stephani \cite{five,twfive} proposed a more
compact description of KS fields and rederived the results of Herlt, giving
also an example of a non-axially symmetric solution. This approach has been
developed further by the present author \cite{twsix,twseven,tweight,twnine}.

In this paper we use the same approach to study fields with vanishing news
function and consequently, with no gravitational radiation. The KS metrics
form a large class of such solutions but do not exhaust all of them. The
intersection between KS and RT fields is represented by the Kinnersley
photon rocket, which explains why more general RT solutions always radiate
gravitationally.

In Section II the notation for twisting algebraically special metrics is
fixed and the expressions for the Bondi mass and the news function are given
according to \cite{twone}. In Section III the condition for no news is
discussed and different classes of metrics that satisfy it are given.

\section{The Bondi mass and news function}

Pure radiation (aligned case) and vacuum algebraically special fields
satisfy the Einstein equations 
\begin{equation}
R_{\mu \nu }=\Phi k_\mu k_\nu  \label{one}
\end{equation}
where $k^\mu $ is the multiple null eigenvector of the Weyl tensor and $\Phi
\geq 0$, vanishing for vacuum solutions. In some coordinates $u.r,\xi ,\bar 
\xi $ the interval reads \cite{five} 
\begin{equation}
ds^2=2\omega \left( dr+Wd\xi +\bar Wd\bar \xi +H\omega \right) -2\frac{%
r^2+\Sigma ^2}{P^2}d\xi d\bar \xi  \label{two}
\end{equation}
where

\begin{equation}
\omega =du+Ld\xi +\bar Ld\bar \xi ,\qquad \partial =\partial _\xi -L\partial
_u  \label{three}
\end{equation}
\begin{equation}
W=-\left( r+i\Sigma \right) L_{,u}+i\partial \Sigma ,\qquad \Sigma =\frac i2%
P^2\left( \partial \bar L-\bar \partial L\right)  \label{four}
\end{equation}
\begin{equation}
H=-r\left( \ln P\right) _{,u}-\frac{mr+M\Sigma }{r^2+\Sigma ^2}+P^2Re\left[
\partial \left( \bar \partial \ln P-\bar L_{,u}\right) \right]  \label{five}
\end{equation}
The functions $P,m,M$ are real and $L$ is a complex function of $u,\xi ,\bar 
\xi $. The Einstein equations read \cite{five}, Eqs.(30.40)-(30.42) 
\begin{equation}
\partial \left( m+iM\right) =3\left( m+iM\right) L_{,u}  \label{six}
\end{equation}
\begin{equation}
M=P^3Im\partial \partial \bar \partial \bar \partial V  \label{seven}
\end{equation}
\begin{equation}
P^3\left[ P^{-3}\left( m+iM\right) \right] _{,u}=P^4\left( \partial
+2\partial \ln P-2L_{,u}\right) \partial I-\frac{\eta ^2}2  \label{eight}
\end{equation}
where $P\equiv V_{,u}$ and 
\begin{equation}
\Phi =\eta ^2\rho \bar \rho ,\qquad \rho ^{-1}=-\left( r+i\Sigma \right)
\label{nine}
\end{equation}
\begin{equation}
I=P^{-1}\left( \bar \partial \bar \partial V\right) _{,u}=\bar \partial
\left( \bar \partial \ln P-\bar L_{,u}\right) +\left( \bar \partial \ln P-%
\bar L_{,u}\right) ^2  \label{ten}
\end{equation}
Eq. (8) is, in fact, a definition of $\eta ^2$, the actual field equations
being (6,7), which are the same as in the vacuum case. Eqs. (7,8) give also 
\begin{equation}
P^3\left( P^{-3}m-Re\partial \partial \bar \partial \bar \partial V\right)
_{,u}=-P^2\left( \partial \partial V\right) _{,u}\left( \bar \partial \bar 
\partial V\right) _{,u}-\frac{\eta ^2}2  \label{eleven}
\end{equation}
where $M$ is excluded and this condition may replace Eq. (8).

In \cite{twone} it was shown that $\bar I$ is the Bondi news function, which
enters the tensor $n_{ab}$ and its derivative 
\begin{equation}
n_{ab}dx^adx^b=-2P_s^{-1}\left( \partial \partial Vd\xi ^2+\bar \partial 
\bar \partial Vd\bar \xi ^2\right)  \label{twelve}
\end{equation}
\begin{equation}
\left( n_{ab}\right) _{,u}dx^adx^b=-2\bar Id\xi ^2-2Id\bar \xi ^2
\label{thirt}
\end{equation}
while the Bondi mass is given by 
\begin{equation}
M_B=m\hat P^{-3}-\frac 12\hat P^{-1}\left( \sigma ^0\bar \sigma ^0\right)
_{,u}+P_s^4Re\left[ \hat P^{-1}I_{,u}\left( \partial \hat u\right)
^2-2\partial I\partial \hat u\right]  \label{fourt}
\end{equation}
where 
\begin{equation}
\hat P\equiv PP_s^{-1},\qquad P_s=1+\frac 12\xi \bar \xi  \label{fift}
\end{equation}
\begin{equation}
\sigma ^0=P_s\partial \partial V,\qquad \left( \sigma ^0\right) _{,u}=P_sP%
\bar I  \label{sixt}
\end{equation}
and $\hat u$ is an approximate Bondi coordinate. Then Eq. (11) becomes the
energy loss formula in the Bondi-Sachs approach.

\section{Fields with no news}

The condition for vanishing news is obviously $I=0$. Eq. (10) gives in this
case 
\begin{equation}
\partial \partial V=F\left( \xi ,\bar \xi \right)   \label{sevent}
\end{equation}
where $F$ is an arbitrary complex function. All, but the first term in the
expression (14) for the Bondi mass, vanish 
\begin{equation}
M_B=\frac{mP_s^3}{P^3}  \label{eightt}
\end{equation}
Eq. (8) simplifies and its real part gives for the mass loss 
\begin{equation}
\left( M_B\right) _{,u}=-\frac{\eta ^2P_s^3}{2P^3}=-\frac{\eta ^2M_B}{2m}
\label{ninet}
\end{equation}
There is no gravitational radiation, the Bondi mass decreases due to the
emission of photons and when $\eta =0$, it stays constant.

A large class of fields, satisfying the no news condition (17) are the
Kerr-Schild fields, which can be defined by \cite{twfive} 
\begin{equation}
\partial \partial V=0  \label{twenty}
\end{equation}
In this case the tensor $n_{ab}$ vanishes identically. One can choose a
gauge with $m=1,M=0,L=ih_\xi $, where $h$ is real. Eq. (19) becomes 
\begin{equation}
\left( \ln V_{,u}\right) _{,u}=\frac{\eta ^2}6  \label{twone}
\end{equation}
which is, in fact, Eq. (32.79) from \cite{five}. One can integrate it twice
to find $V$ as a function of $\eta $. In the vacuum case $V$ must be linear
in $u$ and the reality condition then uniquely fixes $h$. These solutions
have been given already in \cite{three} and include the Kerr black hole. In
the pure radiation case the general solution is not known. Herlt \cite
{five,twfive} has found the general axisymmetric solution. This includes the
radiating Kerr metric \cite{twenty}. An example of a non-axisymmetric
solution was given in \cite{twfive}. In all these metrics $\eta =\eta \left(
u\right) $, so presumably it is more natural to base the classification of
KS fields on the form of $\eta $ and not on the symmetries of the metric.

For RT solutions $L=0$, $M=0$ and the condition $I=0$ becomes 
\begin{equation}
P_{\xi \xi }=0,\qquad P_{\bar \xi \bar \xi }=0  \label{twtwo}
\end{equation}
Together they give the non-axisymmetric solution 
\begin{equation}
P=\alpha \left( u\right) \xi \bar \xi +\beta \left( u\right) \xi +\bar \beta
\left( u\right) \bar \xi +\delta \left( u\right)   \label{twthree}
\end{equation}
where $\alpha ,\delta $ are real. In general, $\alpha ,\beta ,\delta $ may
be interpreted in terms of the acceleration of a particle moving along a
spacelike, timelike or null world line \cite{five}, Sec. 28.3. When $%
K=2\left( \alpha \delta -\beta \bar \beta \right) $ is positive we get the
Kinnersley photon rocket. The case with no $u$-dependence represents the
Vaidya shining star \cite{one}. Consequently, the only RT solution with no
gravitational radiation is the photon rocket, which belongs also to the KS
class. This confirms the previous results \cite{six,nine,thirt,sevent,eightt}%
.

Some general classes of twisting vacuum solutions have been found \cite
{five,thirty} under the assumptions that $L_{,u}-\partial \ln P$ and $m+iM$
are independent of $u$. Among them is the case with no news $I=0$ \cite{five}%
, Sec.29.2.4, which is solved explicitly. Thirty years ago it has been shown 
\cite{thone} that these are the only solutions which are non-radiative in
the sense that the Weyl tensor for large $r$ behaves like $1/r^3$. If the
further condition $L_{,u}=0$ is imposed, the metric becomes independent of $u
$ \cite{five}, Sec.29.2.5. Then $P$ is given by Eq. (23) with constant $%
\alpha ,\beta ,\delta $ and the expressions for $m,M,L$ are given by Eq.
(29.61) from \cite{five}. The subclass $m+iM=const$ contains well-known type
D solutions, such as Kerr and NUT, Kerr and Debney / Demianski
four-parameter solution \cite{thtwo,ththree}. When, in addition, $M=0$ we
come again to the KS vacuum solutions \cite{five}.

\end{document}